\documentclass{appolb}
\usepackage{graphicx}
\usepackage{amsmath,amsfonts,amssymb,hepnames}

\begin{document}
\eqsec  
\title{Entanglement, fluctuations and discrete symmetries in particle decays %
\thanks{Presented at the Workshop on {\it Discrete Symmetries and Entanglement}, Jagiellonian University, Cracow, Poland, 9-11 June 2017}
}
\author{Wojciech Wi\'slicki
\address{National Centre for Nuclear Research, Warsaw, Poland}
}
\maketitle
\begin{abstract}
Pairs of pseudoscalar neutral mesons from decays of vector resonances are studied as bipartite systems in the framework of density operator.
Time-dependent quantum entanglement is quantified in terms of the entanglement entropy and these dependences are demonstrated on data on correlated pairs of $\PK$ and $\PB$ mesons, as measured by the KLOE and Belle experiments .
Another interesting characteristics of such bipartite systems are moments of the CP distributions.
These moments are directly measurable and they appear to be very sensitive to the initial degree of entanglement of a pair.
\end{abstract}
  
\section{Introduction}
Interferometry of neutral mesons is recognized as a powerful and sensitive tool for testing fundamentals of the quantum mechanics.
In particular, pairs of pseudoscalar mesons, originating from strong decays of vector resonances, were proved to be particularly useful in many precision measurements due to their well-defined initial state's quantum numbers and relatively high production rates.
The decays $\Pphi(1020)\rightarrow \PKl\PKs$, $\Ppsi(3770)\rightarrow \PDzero\APDzero$, $\PUpsilon(10580)\rightarrow \PBd\APBd$ and $\PUpsilon(10860)\rightarrow \PBs\APBs$ were recognized long time ago as very useful for the  study of the CP violation \cite{branco}, validity of the CPT and Lorentz invariance \cite{cpt} or search for quantum decoherence \cite{frascati}.
On the other hand, in case of complex initial states where the meson source does neither have well-defined quantum numbers nor symmetry properties and can be a spatially extended object, as e.g. in nuclear collisions, meson interferometry is often used to study both the spatio-temporal characteristics and the degree of coherence of the source.
In any case, either the resonance with well-defined symmetry or a nuclear fireball, the meson pair can be considered as a quantum bipartite system with an arbitrary degree of entanglement in the initial state.
In order to describe the entanglement and its dynamics, a formalism using the reduced density matrix and the entanglement entropy can be easily incorporated and applied to experimental data.
This approach provides a measure of the entanglement in course of the time evolution and an interesting insight into its dependence on the CP, oscillation frequency or the degree of the initial-state entanglement.

Reduced density matrices and entropic approach to bipartite systems show a history of being applied to quantify quantum correlations in condensed matter physics \cite{song}. 
In case of pairs of pseudoscalar mesons, many aspects of quantum entanglement, although not the entropic measures, are a subject of interest, e.g. in the context of decoherence \cite{frascati,decoh} or testing the validity or violation of the CP, T or CPT symmetries \cite{othercpt}.  

Another aspect of interrelations between the CP, entanglement and decay dynamics can be studied by treating the CP of one part of the bipartite system as a random variable and investigating its properties by measuring its moments: the mean value, variance etc.
It turns out that the time-dependent moments of CP, both their absolute values and shapes, are very sensitive to the initial-state entanglement and the decay dynamics of mesons. 
Such approach is as yet unknown in the literature.

\section{Density matrix and entanglement entropy for bipartite systems}
Consider two entangled subsystems A and B in states $|\psi_{A,B}\rangle$ such that the whole bipartite system is in the state $|\psi\rangle=|\psi_A\rangle\otimes |\psi_B\rangle$.
Each subsystem evolves according to its own Hamiltonian acting only in its subspace, $|\psi_{A,B}(t_{A,B})\rangle=\exp(-iH_{A,B}t_{A,B})|\psi_{A,B}\rangle$, where $t_A$ and $t_B$ are the proper times of evolution of the subsystems A and B.  
Initially
\begin{eqnarray}
|\psi(t_A=t_B=0)\rangle=\sqrt{\alpha}|\psi_A\rangle|\psi_B\rangle+\sqrt{1-\alpha}|\psi_B\rangle|\psi_A\rangle,
\end{eqnarray}
where $0\le\alpha\le 1$ parametrizes an initial degree of entanglement between subsystems A and B.
Using the state $|\psi(t_A,t_B)\rangle$, depending on two proper times and represented in any orthonormal basis, one defines the density operator $\rho(t_A,t_B)=|\psi(t_A,t_B)\rangle\langle\psi(t_A,t_B)|$ and the von Neumann entropy
\begin{eqnarray}
S(t_A,t_B)=-\mbox{Tr}\,\rho(t_A,t_B)\ln \rho(t_A,t_B).
\label{eq2.9}
\end{eqnarray} 
Evolution of the density operator in time of any of the subsystems, $t_A$ or $t_B$, is represented by a unitary transformation defined by the subsystem's Hamiltonian and acting only in its appropriate subspace:
\begin{eqnarray}
\rho(t_A,\,.\,) & = & e^{-iH_At_A}\,\rho(0,\,.\,)\,e^{iH_At_A} \nonumber \\
\rho(\,.\,,t_B) & = & e^{-iH_Bt_B}\,\rho(\,.\,,0)\,e^{iH_Bt_B}.
\label{eq2.8}
\end{eqnarray}
Since (\ref{eq2.8}) represents unitary transformations, so $\rho(t_A,t_B)$ has all required properties of the density operator: Hermiticity, positivity and unit trace. 
Moreover, it has the same von Neumann entropy (\ref{eq2.9}) as $\rho(0,0)$.
Here the time evolution is naturally determined by the subsystems' dynamics.
This opens a way to test some possible extensions of the physical picture, e.g. by adding the non-Standard Model or non-Hermitean terms to $H_{A,B}$ and looking for their effects on the time evolution, modifications to meson's lifetimes or masses, etc.

By using the reduced density operator $\rho_A=\mbox{Tr}_B\rho$, obtained by tracing over the degrees of freedom of the subsystem B and integration over $t_B$, one defines the entanglement entropy
\begin{eqnarray}
S_A(t_A)=-\mbox{Tr}\,\rho_A(t_A)\ln \rho_A(t_A).
\label{eq2.3}
\end{eqnarray}
Since $S\le S_A+S_B$ (equality for unentangled subsystems) and $S_A=S_B$, the mutual information $I$
\begin{eqnarray}
I(\rho) & = & S(\rho_A)+S(\rho_B)-S(\rho) \nonumber \\
        & = & 2S_A-S
\end{eqnarray}
quantifies the entanglement between subsystems A and B.

\section{Degree of entanglement from fits to time-dependent decay spectra of pairs of neutral mesons}
Availability of experimental data motivates us to perform calculations for two decays: $\Pphi(1020)\rightarrow \PKl\PKs$ \cite{kloe} and $\PUpsilon(4S)=\PUpsilon(10580)\rightarrow \PBd\APBd$ \cite{belle}.
Flavour eigenstates of the final-state mesons exhibit a particle-antiparticle mixing due to the weak box processess \cite{branco}.
We consider these decays in the rest frame of the initial resonance but decay times are measured each in the rest frame of the decaying neutral meson.
Final-state mesons fly back-to-back with identical momenta and are detected using their decays in two detectors. 
For simplicity, we assume that both mesons decay to the same final state.

\begin{figure}[h]
\centerline{
\includegraphics[scale=.35]{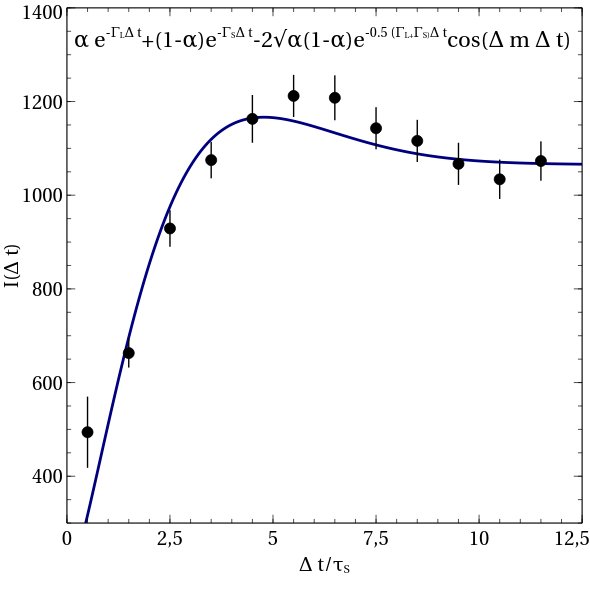} \hspace{5mm} \includegraphics[scale=.35]{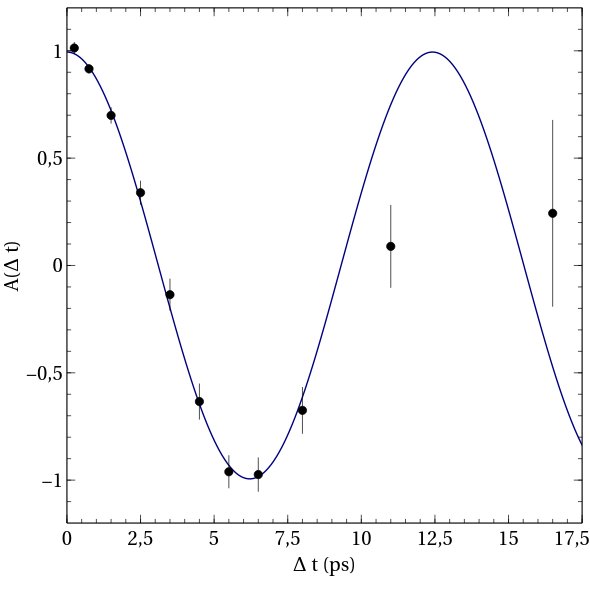}}
\caption{Left: Intensity spectrum of decay time difference of pairs $\PKl,\PKs\rightarrow\Ppiplus\Ppiminus$ measured by KLOE \cite{kloe} with a fit of eq. (\ref{eq3.2}); Right: Asymmetry between unmixed and mixed final states in semileptonic decays of $\PBz, \PaBz$, as measured by Belle \cite{belle} with a fit of eq. (\ref{eq3.3}). In both panels, the statistical and systematic errors were combined.} 
\label{fig1}
\end{figure}

For $\Pphi(1020)\rightarrow \PKl\PKs$, where $J^{PC}(\Pphi)=1^{--}$, the final state has to be antisymmetric and at the moment of decay it reads
\begin{eqnarray}
|\psi(t_A=0,t_B=0)\rangle = 2^{-1/2}(\sqrt{\alpha}|\PKl\rangle_{A}|\PKs\rangle_{B}-\sqrt{1-\alpha}|\PKs\rangle_{A}|\PKl\rangle_{B}),
\end{eqnarray}
where subcripts $_{A,B}$ refer to the detectors.
We consider only decays to the same final states $\Ppiplus\Ppiminus$.
The decay intensity dependence on $\Delta t=|t_B-t_A|$, after integrating over $t_A+t_B$, reads
\begin{eqnarray}
I(\Delta t)=\alpha e^{-\Gamma_L\Delta t}+(1-\alpha)e^{-\Gamma_S\Delta t}-2\sqrt{\alpha(1-\alpha)}e^{-\bar\Gamma\Delta t}\cos(\Delta m\Delta t)
\label{eq3.2}
\end{eqnarray}
where $\Gamma_{L,S}$ stand for decay rates of $\PKl,\PKs$, $\bar\Gamma=(\Gamma_L+\Gamma_S)/2$ and $\Delta m =m_L-m_S\sim 1\,\mbox{ns}^{-1}$.
The long- and shortliving kaons $\PKl$ and $\PKs$ were experimentally identified by their decay times.
Fitting eq. (\ref{eq3.2}) to the decay spectrum measured by KLOE \cite{kloefp} one gets $\alpha=0.71\pm 0.31$ (cf. fig. \ref{fig1}).

In the similar case of $\PUpsilonFourS\rightarrow \PBd\APBd$, the antisymmetric final state parametrized with the degree of entanglement $\alpha$ is initially (neglecting small CP-violation effect) equal to
\begin{eqnarray}
|\psi(t_A=t_B=0)\rangle & = & 2^{-1/2}(\sqrt{\alpha}|\PB_H\rangle_{A}|\PB_L\rangle_{B}-\sqrt{1-\alpha}|\PB_{L}\rangle_{A}|\PB_H\rangle_{B}) \nonumber \\
                        & = & 2^{-1/2}(\sqrt{\alpha}|\PBz\rangle_{A}|\PaBz\rangle_{B}-\sqrt{1-\alpha}|\PaBz\rangle_{A}|\PBz\rangle_{B})
\end{eqnarray} 
and from its time-dependent states we build up the time-dependent asymmetry between the flavour-unmixed and mixed states \cite{belle}
\begin{eqnarray}
A(\Delta t) & = & \frac{N_u(\Delta t)-N_m(\Delta t)}{N_u(\Delta t)+N_m(\Delta t)} \nonumber \\
            & = & \frac{2\sqrt{\alpha(1-\alpha)}\cos(\Delta m\Delta t)}{\alpha e^{-\Delta\Gamma\Delta t/2}+(1-\alpha)e^{\Delta\Gamma\Delta t/2}},
            \label{eq3.3}
\end{eqnarray}
where $\Delta\Gamma=\Gamma_H-\Gamma_L$ is a small number consistent with zero within experimental errors and $\Delta m=m_H-m_L=0.506\,\mbox{ps}^{-1}$.  
The mixed and unmixed states are identified using the lepton sign of the decay $\PBz(\PaBz)\rightarrow \PB^{(\ast)-(+)}\Pmu^{+(-)}\Pnum(\APnum)$.
By fitting eq. (\ref{eq3.3}) to the Belle data \cite{belle} (cf. Fig. \ref{fig1}) one obtains $\alpha=0.55\pm 0.07$. 

In both cases, the values of $\alpha$ from fits to data do not indicate any significant deviation from the maximal entanglement $\alpha=1/2$. 
Important to note, the non-maximal entanglement would violate antisymmetry of the final state.
That could indicate an ill-defined CPT operator and a particle-antiparticle identity, and lead to exotic and intriguing consequences \cite{bernabeu}. 

\section{Entanglement entropies for pairs of neutral mesons}
The density operator $\rho(t_A,t_B)$ has to be expressed in an orthonormal basis.
In case of neutral mesons, the orthonormal basis $(|\PK_1\rangle,|\PK_2\rangle )$ differs from the non-orthonormal one $(|\PKl\rangle,|\PKs\rangle )$ due to the CP violation, parametrized by a complex parameter $\varepsilon$, where $|\varepsilon |=2.2\times 10^{-3}$ and $\phi_{\varepsilon}=43.5^{\circ}$ are precisely known from experiment.
Direct CP violation effects are much smaller and are neglected.
For neutral $\PB$ mesons, the CP violation effect is smaller than $10^{-3}$ and can be neglected so that the basis $(|\PB_H\rangle, |\PB_L\rangle )$ is considered to be orthonormal.

For kaons, the matrix elements of the reduced density matrix $\rho_{A_{i,j}}=\linebreak _B\langle \PK_{1,2}|\rho |\PK_{1,2}\rangle_B$, i.e. after tracing over degrees of freedom of the meson in detector B, and after integrating over $t_B$ and to the order ${\mathcal O}(\varepsilon)$, read
\begin{eqnarray}
\rho_{A_{11}}(t) & = & \frac{\alpha}{\Gamma_L}e^{-\Gamma_St} \nonumber \\
\rho_{A_{22}}(t) & = & \frac{1-\alpha}{\Gamma_S}e^{-\Gamma_Lt} \nonumber \\
\rho_{A_{12}}(t) & = & \frac{\varepsilon^{\ast}\alpha}{\Gamma_L}e^{-\Gamma_St} + \frac{\varepsilon(1-\alpha)}{\Gamma_S}e^{-\Gamma_Lt} \nonumber \\
                 & + & \frac{2(\Re\epsilon)\sqrt{\alpha(1-\alpha)}}{\bar\Gamma^2+(\Delta m)^2}e^{-\bar\Gamma t}(\bar\Gamma e^{i(\Delta m)t}+(\Delta m)e^{i(\Delta m-\pi/2)t}) \nonumber \\
                 \rho_{A_{21}}(t) & = & \rho_{A_{12}}^\ast(t).
\label{eq4.1}
\end{eqnarray} 
Since the mesons decay, the density operator has to be renormalized $\rho_A(t)\rightarrow \rho_A(t)/\mbox{Tr}\rho_A(t)$ in order to meet the normalization requirement $\mbox{Tr}\rho_A(t)=1$.

Using eq.(\ref{eq4.1}), the entanglement entropy (\ref{eq2.3}) is found to be (t-dependence omitted for simplicity)
\begin{eqnarray}
S_A & = & -\rho_{A_{11}}\ln\rho_{A_{11}}-\rho_{A_{22}}\ln\rho_{A_{22}} \nonumber \\
    & - & 2[(\Re\,\rho_{A_{12}})\ln |\rho_{A_{12}}|+(\Im\,\rho_{A_{12}})\arg \rho_{A_{12}}].
\label{eq4.2}
\end{eqnarray}
Similar formulae to eqns (\ref{eq4.1}) and (\ref{eq4.2}) can be found for $\PB$ mesons.

\begin{figure}[h]
\centerline{
\includegraphics[scale=.33]{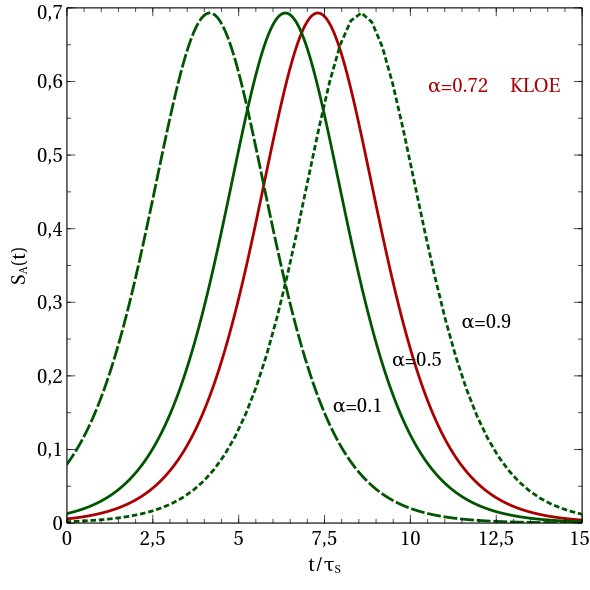} \hspace{8mm} \includegraphics[scale=.33]{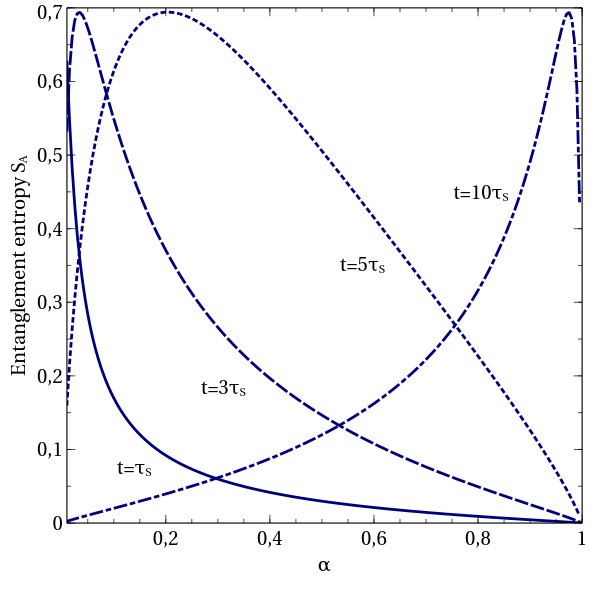}}
\centerline{
\includegraphics[scale=.33]{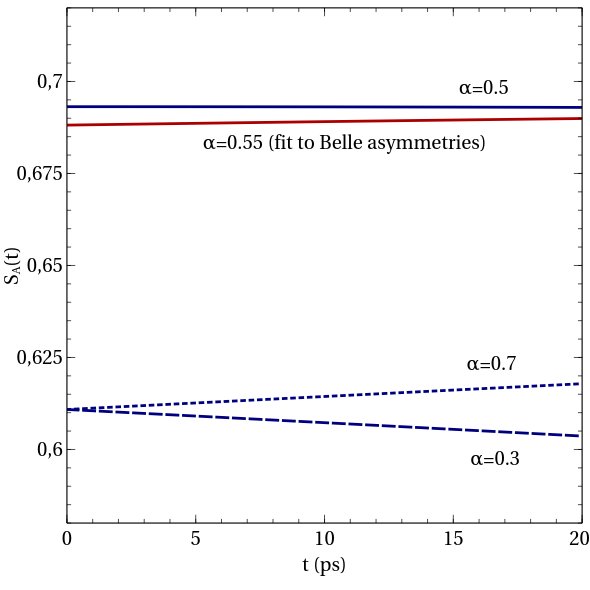} \hspace{8mm} \includegraphics[scale=.33]{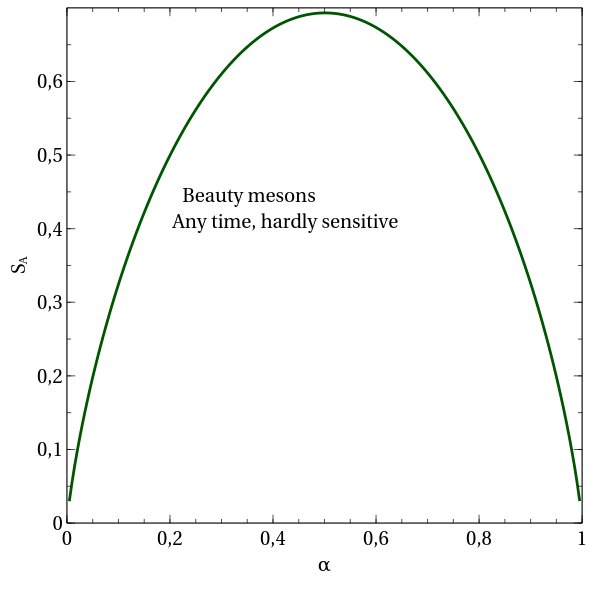}}
\caption{Upper left: Time evolution of the entanglement entropy $S_A$ for neutral $\PK$'s, for a number of entanglement parameters $\alpha$; Upper right: $\alpha$-dependence of $S_A$ for a number of times; Lower left and right: the same dependencies for neutral $\PB$'s.} 
\label{fig2}
\end{figure}
Fig. \ref{fig2} presents the entanglement entropy dependence on time and the initial entanglement degree $\alpha$.
For $\PK$ mesons, the entanglement entropy exhibits interesting dependence on $\alpha$, and a time-dependence governed by the oscillation frequency $\Delta m$ and large lifetime difference between $\PKs$ and $\PKl$. 
The time when the entanglement is maximal during evolution strongly depends on $\alpha$ and is related to the location of the interference maximum.
Contrary to kaons, the initial value of $S_A$ for pairs of $\PB$'s is very sensitive to $\alpha$ but its later time dependence is weak, due to much smaller difference of lifetimes of $\PB_H$ and $\PB_L$. 
However, for given $\alpha$ an average entanglement entropy in later times for $\PB$'s is larger compared to $\PK$'s.
Dependence of $S_A$ on $\alpha$ for $\PB$'s is almost the same for all times.

\section{Fluctuations of CP}
Mesons in pairs originating from decays of vector resonances carry opposite CP.
Since the direction of emission of a meson with given CP is purely random, the CP at given detector could be naively expected to be a simple, binary random variable.
But quantum entanglement, the dynamics of time evolution and the initial degree of entanglement make this observable less obvious and more intriguing.  
We show here some of its interesting dependencies and argue that the moments of CP are sensitive probes of the initial entanglement of a pair $\alpha$.
Unlike the entanglement entropy, it does not quantify the time-dependent entanglement itself but the CP registered by one detector which, contrary to the naive expectation, exhibits a highly non-trivial and $\alpha$-sensitive time-dependence.

Moments of CP can be found using the time-dependent moment generating function $\chi(\lambda,t)$ and differentiating it to obtain cummulants $C_n,\;,n=1,2,\ldots $
\begin{eqnarray}
\chi(\lambda,t) & = & \langle \exp (i\lambda\cdot \mbox{CP}_A)\rangle \nonumber \\
                & = & \sum_m P(CP_A=m) e^{i\lambda m}, \nonumber \\
\nonumber \\
C_n & = & \big (-i\frac{\partial}{\partial\lambda}\big )^n\ln\chi(\lambda,y)|_{\lambda=0},
\end{eqnarray}
where $\langle \ldots\rangle$ stands for the expected value in the state $|\psi(t)\rangle$ and $\mbox{CP}_A$ is the CP registered in detector A. 
The first moments $C_1$ and $C_2$ correspond to the expected value and variance, respectively.
Noteworthy, these moments are directly measurable by registering the identified long- or short-living mesons in one of detectors and correcting for CP-violation.

\begin{figure}[h]
\centerline{
\includegraphics[scale=.35]{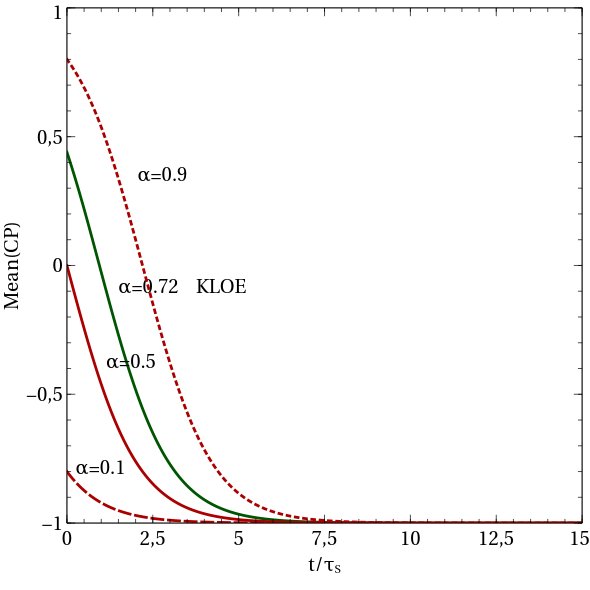} \hspace{5mm} \includegraphics[scale=.35]{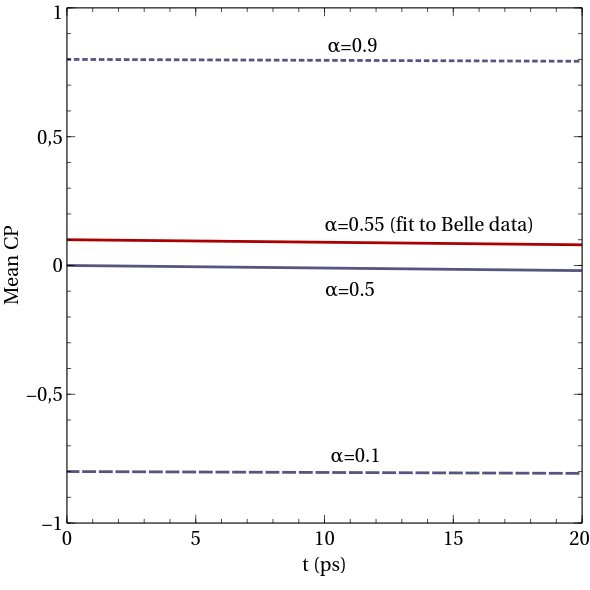}}
\caption{Left: the time dependence of the mean CP in detector A for neutral $\PK$ meson pairs, parametrized by the degree of initial coherence $\alpha$; Right: the same for the $\PB$ meson pairs.} 
\label{fig3}
\end{figure}

For the $\Pphi(1020)\rightarrow \PKl\PKs$, keeping only terms to linear order in $\varepsilon$, the cummulant generating function is equal to
\begin{eqnarray}
\chi(\lambda,t) & = & P(\mbox{CP}_A=+1)e^{i\lambda}+ P(\mbox{CP}_A=-1)e^{-i\lambda} \nonumber \\
                & = & \frac{\alpha e^{-\Gamma_St}e^{-i\lambda}}{\Gamma_L^2/4+m_L^2} +\frac{(1-\alpha) e^{-\Gamma_Lt}e^{i\lambda}}{\Gamma_S^2/4+m_S^2}.
\end{eqnarray}
The first moment, or the mean value, of CP is equal to
\begin{eqnarray}
C_1(t,\alpha) & = & \langle \mbox{CP}_A(t)\rangle \nonumber \\
              & = & \frac{\alpha-(1-\alpha)\frac{\Gamma_L^2+4m_L^2}{\Gamma_S^2+4m_S^2}e^{\Delta\Gamma t}}{\alpha+(1-\alpha)\frac{\Gamma_L^2+4m_L^2}{\Gamma_S^2+4m_S^2}e^{\Delta\Gamma t}}.
\label{eq5.4}
\end{eqnarray}
In particular
\begin{eqnarray}
C_1(t,\alpha) & \xrightarrow[t\rightarrow 0]{} & 2\alpha -1 \nonumber \\
C_1(t,\alpha) & \xrightarrow[t\rightarrow \infty]{} & -1.
\end{eqnarray}
Similar formulae are valid for $\PB$ mesons.
In the limit of long time, only the long-living components with CP=-1 survive and all short-living ones with CP=+1 die out.
This effect is clearly seen for $\PK$ mesons where the difference between the $\PK_1$ and $\PK_2$ lifetimes is large.
For the $\PB$ mesons, the time dependence is qualitatively the same but weaker since the $\PB_H$ and $\PB_L$ lifetimes are close to each other. 
Fig. \ref{fig3} presents the $C_1$ time-dependence for the $\PK$ and $\PB$ pairs. 

The second moment, or the variance of CP, is equal to
\begin{eqnarray}
C_2(t,\alpha) & = & \langle \mbox{CP}_A^2\rangle - \langle \mbox{CP}_A\rangle ^2 \nonumber \\
              & = & 1 - \bigg[\frac{\alpha-(1-\alpha)\frac{\Gamma_L^2+4m_L^2}{\Gamma_S^2+4m_S^2}e^{\Delta\Gamma t}}{\alpha+(1-\alpha)\frac{\Gamma_L^2+4m_L^2}{\Gamma_S^2+4m_S^2}e^{\Delta\Gamma t}}\bigg]^2.
\end{eqnarray}
and similar formulae hold for $\PB$ mesons.
Particular values are
\begin{eqnarray}
C_2(t,\alpha) & \xrightarrow[t\rightarrow 0]{} & 4\alpha (1-\alpha) \nonumber \\
C_2(t,\alpha) & \xrightarrow[t\rightarrow \infty]{} & 0.
\end{eqnarray}
For large values of $t$ only the long-living components $\PKl$ and $\PB_H$ survive and the CP distribution narrows down to zero width. 
\begin{figure}[h]
\centerline{
\includegraphics[scale=.35]{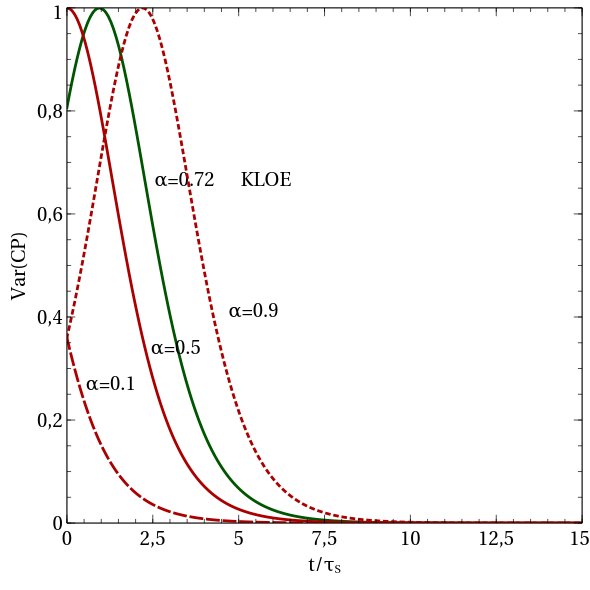} \hspace{5mm} \includegraphics[scale=.35]{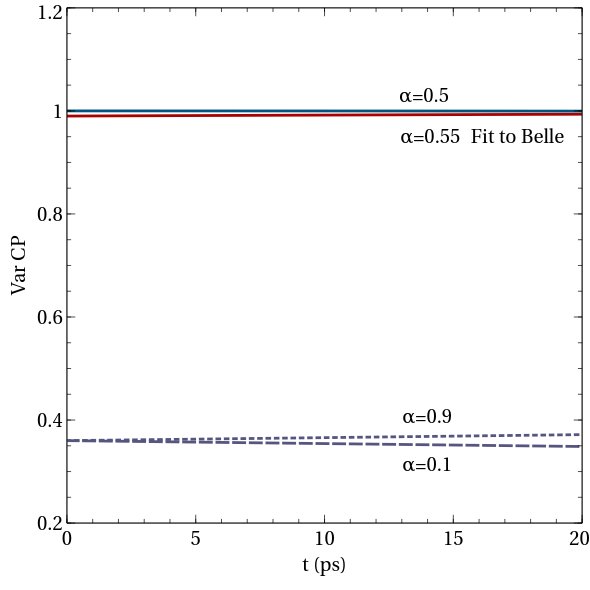}}
\caption{Left: the time dependence of the variance of CP in detector A for neutral $\PK$ meson pairs, parametrized by the degree of initial coherence $\alpha$; Right: the same for the $\PB$ meson pairs.} 
\label{fig4}
\end{figure}
The variance can be non-monotonic only for kaons and the maximum is located at $t_0=\frac{1}{\Delta \Gamma}\ln\big[\alpha(1-\alpha)\frac{\Gamma_S^2+4m_S^2}{\Gamma_L^2+4m_L^2}\big]$ and $t_0>0$ only for $\alpha\gtrsim 0.5$. 
The $t_0$ becomes infinite as $\alpha\rightarrow 1$. 

\section{Conclusions}
It this paper we propose an approach providing a new insight into the quantum entanglement in neutral meson pairs.
Such pair is treated as a quantum bipartite system where the degree of initial entanglement is allowed to be a free parameter, thus allowing for a possible imperfect coherence in the initial decay and quark hadronization. 
At the same time, it also allows to test the parity of the final state and thus examine the correctness of the fundamental assumptions on the Bose-Einstein symmetry and CPT invariance.
The entanglement parameter was determined from data on decays $\Pphi(1020)\rightarrow \PKl\PKs$ and $\PUpsilon(4S)\rightarrow \PBd\APBd$ and found to be consistent with maximal entanglement, although with a rather large error.
In order to quantify the degree of entanglement in course of the time evolution of a system, the entanglement entropy is calculated and discussed. 
This quantity appears to be very sensitive to the initial entanglement and exhibits interesting dynamics.
Another quantity, strongly dependent on both the entanglement and dynamics of meson decays, are moments of the CP of one subsystem. 
These observables exhibit stronger and non-monotonic time dependence for pairs of $\PK$ mesons than for $\PB$ mesons due to larger lifetime difference between the components of opposite CP. 

This work was supported by the NCN grant 2013/08/M/ST2/00323.
 


\begin{thebibliography}{99}
\bibitem{branco} G.C. Branco, L. Lavoura and J.P. Silva, {\it CP Violation}, Clarendon Press, Oxford, 2007
\bibitem{cpt} KLOE-2: D. Babusci et al., {\it Phys. Lett.} B730 (2014) 89, LHCb: R. Aaij et al., {\it Phys. Rev. Lett. 116 (2016) 241601}
\bibitem{frascati} {\it Handbook on Neutral Kaon Interferometry at a $\Phi$-Factory}, ed. A. di Domenico, Frascati Physics Series, vol. XLIII, Frascati, 2007
\bibitem{song} H. Francis Song et al., {\it Phys. Rev.} B85 (2012) 035409
\bibitem{decoh} J. Bernabeu et al., {\it Phys. Rev.} D74 (2006) 045014
\bibitem{othercpt} M. Nebot, {\it J. Phys. Conf. Ser.} 873 (2017) 012024, \\
                    A. di Domenico, {\it Acta Phys. Polon.} A127 (2015) 1563, \\
                    J. Bernabeu and F. Martinez-Vidal, {\it Rev. Mod. Phys.} 87 (2015) 165, \\
                    Zhije Huang and Yu Shi, {\it Phys. Rev.} D89 (2014) no.1, 016018 
\bibitem{kloe} KLOE: F. Ambrosino et al., {\it Phys. Lett.} B636 (2006) 173
\bibitem{belle} Belle: A. Go et al., {\it Phys. Rev. Lett.} 99 (2007) 131802
\bibitem{kloefp} KLOE: A. di Domenico et al., {\it Found. Phys.} 40 (2010) 852
\bibitem{bernabeu} J. Bernabeu et al., {\it Nucl. Phys.} B744 (2006) 180 
\end{thebibliography}
\end{document}